\documentstyle[preprint,aps]{revtex}
\tightenlines

\newcommand{\ket}[1]{|{#1}\rangle}
\newcommand{\bra}[1]{\langle{#1}|}

\begin{document}

\title{Interferometric Tests of Teleportation}
\author{T.C.Ralph}
\address{Department of Physics, Centre for Lasers, 
\\ The University of Queensland, \\ St Lucia 4072 Australia \\ 
 E-mail: ralph@physics.uq.edu.au}
\maketitle

\begin{center}
\scriptsize (December 2000)
\end{center}

\begin{abstract}

We investigate a direct test of teleportation efficacy based on a 
Mach-Zehnder interferometer. The 
analysis is performed for continuous variable teleportation of both 
discrete and continuous observables.  

\end{abstract}

\vspace{10 mm}

\section{Introduction}

Information is not independent of the physical laws used to store and process it 
\cite{lan91}. The unique properties of quantum mechanics lead to 
radically different ways 
of communicating and processing information \cite{ben95}. The study of 
Òquantum informationÓ is currently one of the fastest growing areas of physics.

Quantum teleportation \cite{ben93,bou97,bos98,pan98,fur98} 
is a method by which quantum 
information can be passed through a classical channel and successfully 
retrieved at a distant location. The sharing of entanglement between 
the sender (Alice) and receiver (Bob) is essential for 
teleportation as it provides the ``quantum key'' needed to retrieve 
the quantum information \cite{deu99}. 
In this way an unknown quantum state of an object 
can be transferred through a classical channel, 
with neither Bob nor Alice knowing the state. 

The efficacy of teleportation can be characterized in two quite 
distinct ways. Traditionally fidelity is used for this purpose 
\cite{sch95}. Fidelity, $F$, gives a measure of the quality of the 
teleported state by evaluating the overlap between the input 
state, $|\psi \rangle$, and 
the teleported output state, $\rho$, via $F=\langle \psi|\rho|\psi \rangle$. 
Fidelity is state dependent, ie the fidelity of the reconstructed state 
depends both on the quality of the teleporter and on the class of 
input states from which the unknown state is picked. More recently 
the amplitude conditional 
variance between the input and output has been suggested as an alternative 
measure \cite{ral98,ral001}. The conditional variance measures the amount 
of uncorrelated noise that is added to the quantum state in the 
teleportation process. As such it is a measure of the quality of the 
teleporter itself, independent of the state to be teleported. 
The measurement of either fidelity or the conditional variance involves a 
``third'' person, Victor (the verifier), who
prepares the input states and examines the teleported states to 
determine the quality of the teleportation. For example 
Victor may prepare photons in particular polarisation states and then 
check if they are still in the same states after teleportation 
\cite{bou97,bos98}. For continuous variable experiments the signal and 
noise properties of the input and output are compared 
\cite{fur98,ral99}.  

Another way of testing the efficacy of teleportation is to create a 
pair of quantum correlated objects, teleport one of them, then test 
directly to what extent they are still quantum correlated. An example 
of this is polarization entanglement swapping \cite{pan98} in which one of an 
entangled pair of photons is teleported and then the degree of 
entanglement that remains between them is measured. A version of this 
experiment using continuous variable teleportation has also been 
proposed \cite{pol99}. Another possibility is to teleport one arm 
of a spatial superposition and then measure the preservation of the 
superposition directly through their interference characteristics. 
These types of tests are important for 3 reasons: (i) They directly 
observe the preservation of quantum correlations rather than just 
inferring them; (ii) such specific situations highlight 
aspects of the physics of the teleportation process not obvious from 
considering more general figures of merit and; (iii) from a practical 
point of view the preservation of entanglement and interference 
effects will be an important aspect of teleportation in most quantum 
processing applications.

A spatial superposition test can be applied to single photon 
polarization states
using a Mach-Zehnder interferometer \cite{ral992}. 
An interesting feature of such a 
test is that it is possible for Alice and Bob to verify 
that their teleporter is operating correctly without knowing the input 
states. In this 
paper we generalize this test to cover a broad range of input states, 
including continuous variable states. We will begin, in section 2, by 
introducing the model for a teleporter we will use throughout the 
paper. In section 3 we will review the operation of the 
Mach-Zehnder interferometer as a teleportation tester for single photon, 
polarisation superposition inputs. Section 4 will examine more general 
low photon number states. In section 5 we will
generalize the technique to input states with continuous degrees of 
freedom and in section 6 we will discuss and conclude.

\section{The Teleporter}

The teleporter we will consider in this paper is an all optical 
device using continuous variable (squeezing) entanglement as a quantum 
resource \cite{ral991}. This model is chosen for its 
versatility in being able to teleport all the input states considered 
in this paper. In an experimental situation more input specialized devices 
may be used. Consider first the ``classical 
teleportation'' device depicted in Fig.1(a). By classical we mean we 
attempt to transfer the quantum information through a classical 
channel without the assistance 
of entanglement. The input light field, $\hat a_{in}(t)$, 
is sent through a linear optical
amplifier by Alice. In Fourier space the output of a linear amplifier 
can be written
\begin{equation}
a_{c}(\omega)=\sqrt{\eta_{a} 
G(\omega)}a_{in}(\omega)+\sqrt{(G(\omega)-1)}v_{1}^{\dagger}+
\sqrt{G(\omega)(1-\eta_{a})}v_{a}
\label{ac}
\end{equation}
where $G(\omega)$ is the (frequency dependent) amplifier gain and 
$v_{1}$ and $v_{a}$ are vacuum noise inputs due to the gain and 
internal losses ($\eta_{a}$) of the amplifier respectively.
If the gain is sufficiently large ($G>>1$) then $a_{c}$ can be
regarded as a classical field. This is because the conjugate quadrature
variables $X_{c}^{+}=a_{c}+a_{c}^{\dagger}$ and
$X_{c}^{-}=i(a_{c}-a_{c}^{\dagger}$) both have uncertainties much
greater than the quantum limit, i.e. $\Delta (X_{c}^{\pm})^{2}>>1$.
This means that simultaneous measurements of the
conjugate quadratures can extract
all the information carried by $a_{c}$ with negligible penalty. The quantum
noise added due to the simultaneous measurements will be negligible
compared to the amplified quadrature uncertainties. It is thus 
possible to convert then transmit the information carried in this beam over any 
available classical channel (radio, copper wires, etc).
However it is convenient, and no less general, 
to retain an optical classical channel. Further discussion and a 
simple proof of the classical nature of this  
channel can be found in the appendix. 

When Bob receives the classical beam he attempts
to retrieve the quantum state of the input by simply attenuating
the beam with a beamsplitter of
transmission $\varepsilon$. The output field is
$a_{out}=\sqrt{\varepsilon}a_{c}-\sqrt{1-\varepsilon}v_{2}$ where 
$v_{2}$ is the
vacuum mode incident on the unused port of the beamsplitter. The 
final output field is thus
\begin{equation}
a_{out}(\omega)= \lambda(\omega) a_{in}(\omega)+
({{\lambda(\omega)}\over{\sqrt{\eta_{a}}}} 
v_{1}^{\dagger}-v_{2})+\lambda(\omega){{\sqrt{1-\eta_{a}}}
\over{\sqrt{\eta_{a}}}}v_{a}
\label{sn}
\end{equation}
where the total classical channel gain is given by $\lambda(\omega)=
\sqrt{G(\omega) 
\varepsilon \eta_{a}}$ and we have assumed the classical channel 
limit $G\to \infty$ and $\varepsilon \to 0$. 
In practice we are only interested in finite bandwidths. For photon 
counting experiments this usually means frequency filters 
will be placed in front of the detectors. For continuous variable 
experiments only a finite range of RF frequencies will be analyzed. 
We will assume that the optical amplifier, and thus $\lambda$ has a 
flat response over the detection bandwidth. Hence, setting unity gain 
($\lambda=1$) and negligible loss ($\eta_{a}=1$) 
we obtain the usual result
\begin{equation}
a_{out}=a_{in}+v_{1}^{\dagger}-v_{2}
\end{equation}
whereby two vacuum noise penalties are imposed by classical 
teleportation \cite{bra98,ral98}.

Quantum teleportation can be achieved by replacing the independent 
vacuum inputs, $v_{1}$ and $v_{2}$, with Einstein,
Podolsky, Rosen (EPR) entangled beams \cite{ou92}, $b_{1}$ and 
$b_{2}$, as shown in Fig.1(b). Such beams
have the very strong correlation property that both their difference
amplitude quadrature variance,
$\Delta(X_{b1}^{+}-X_{b2}^{+})^{2}$, and their sum phase quadrature
variance, $\Delta(X_{b1}^{-}+X_{b2}^{-})^{2}$, are less than the
quantum limit (=1). Such beams can be generated by sub-threshold non-degenerate
parametric amplification \cite{ou92} or by the mixing of independent squeezed
sources \cite{yeo93,ral98}. For non-degenerate parametric amplification
these beams can be represented by
\begin{eqnarray}
b_{1}(\omega) & = & 
\sqrt{\eta_{b1}H(\omega)}v_{3}+\sqrt{\eta_{b1}(H(\omega)-1)}v_{4}^{\dagger}+
\sqrt{1-\eta_{b1}}v_{b1}\nonumber\\
b_{2}(\omega) & = & 
\sqrt{\eta_{b2}H(\omega)}v_{4}+\sqrt{\eta_{b2}(H(\omega)-1)}v_{3}^{\dagger}+
\sqrt{1-\eta_{b2}}v_{b2}
\label{sa}
\end{eqnarray}
where $H(\omega)$ is the parametric gain and as before the $\eta$'s and $v$'s are 
efficiencies and resultant vacuum inputs respectively. The strength of 
the squeezing entanglement can be characterized by 
$V_{ent}=(\sqrt{H}-\sqrt{H-1})^{2}$ which varies from not entangled 
($V_{ent}=1$) to strongly entangled ($V_{ent} \to 0$) as the 
parametric gain increases. We will also refer to the percentage of 
entanglement squeezing as $(1-V_{ent})\times 100\%$. The output 
field is now given by
\begin{equation}
a_{out}(\omega)= \lambda(\omega) a_{in}(\omega)+
({{\lambda(\omega)}\over{\sqrt{\eta}}} 
b_{1}^{\dagger}(\omega)-b_{2}(\omega))+
\lambda(\omega){{\sqrt{1-\eta}}\over{\sqrt{\eta}}}v_{a}
\label{ss}
\end{equation}
which, because of the strong correlations between $b_{1}$ and $b_{2}$, 
reduces to
\begin{equation}
a_{out}(\omega) = \lambda (\omega) a_{in}(\omega)+
(\lambda (\omega) \sqrt{H(\omega)}-\sqrt{H(\omega)-1})v_{3}^{\dagger}+
(\sqrt{H(\omega)}-\lambda \sqrt{H(\omega)-1}) v_{4}
\label{tel}
\end{equation}
in the absence of losses ($\eta_{a}=\eta_{b1}=\eta_{b2}=1$). Again we 
assume (and will do so for the remainder of the paper) that all gains 
are flat across the detection bandwidth. 
In the limit of very high parametric gain ($H \to \infty$, 
$V_{ent}\to 0$) and unity 
classical channel gain ($\lambda=1$) the
output becomes identical to the input ($a_{out} \to a_{in}$). This is 
ideal
quantum teleportation as the only direct link between the input and
output is the classical field $a_{c}$, yet arbitrarily accurate reconstruction of
the input state is, in principle, possible with a sufficiently strong EPR
correlation. The uncertainty principle is not compromised because the
variances of each of the quadratures of $b_{1}$ by themselves are very noisy.
Thus the information about $a_{in}$ carried on the classical field is
buried in this noise and cannot be extracted by using the
classical field alone. An important operating point is the classical 
channel gain $\lambda_{opt}={{\sqrt{H-1}}\over{\sqrt{H}}}$. With this 
gain, in the absence of losses, the output field is given by
\begin{equation}
a_{out} = \lambda_{opt} a_{in}+(\sqrt{1-\lambda_{opt}^{2}}) v_{4}
\label{att}
\end{equation}
i.e. it is simply an attenuated version of the input \cite{pol99}. 
The teleporter can be generalized to deal with arbitrary 
polarisations of the input field by decomposing the field into 
orthogonal polarisation components (using a polarising beamsplitter) 
and teleporting the individual components separately (see Fig.1(c)). 

The question remains as to how the linear amplifier in Fig.1 could
be constructed. This is not trivial as in standard optical amplifiers
the source of the vacuum mode is not available for modification.
For example, in a laser amplifier the physical origin
of the vacuum input ($v_{1}$) is collisionally or phonon induced dipole
fluctuations of the gain medium \cite{yam86}. One solution is shown
schematically in Fig.2. The input beam is mixed with the EPR beam,
$b_{1}$, at a 50:50 beamsplitter. The output beams are
\begin{eqnarray}
c & = & {{1}\over{\sqrt{2}}}(a_{in}+b_{1})\nonumber\\
d & = & {{1}\over{\sqrt{2}}}(a_{in}-b_{1})
\end{eqnarray}
The beams are amplified by degenerate parametric
amplifiers of equal gains but with a $\pi$ phase shift between there
pump ($E$) phases. This results in the outputs
\begin{eqnarray}
c' & = & \sqrt{G}c+\sqrt{G-1}c^{\dagger}\nonumber\\
d' & = & \sqrt{G}d-\sqrt{G-1}d^{\dagger}
\end{eqnarray}
Recombining these beams on a beamsplitter then produces the desired
output: $a_{c}=\sqrt{G}a_{in}+\sqrt{G-1}b_{1}^{\dagger}$.

\section{The Mach-Zehnder Interferometer and the Teleporter}

We now examine the efficacy of the teleporter described in the 
previous section as characterized using an interferometer. In this 
section we will consider idealized single photon polarisation 
superpositions as inputs to illustrate the basic 
physics.  
In the next section we will consider more general polarisation-number 
inputs. In the following section continuous 
variable inputs will be considered.

Consider first the set-up shown schematically in Fig.3(a) (see also 
Fig.4(a)). Basically we 
place a teleporter in one arm of a Mach-Zehnder interferometer, inject 
a single photon state, in an arbitrary polarisation superposition 
state into one port, then use the interference visibility at the 
output ports to characterize the efficacy of teleportation. A useful 
feature of this 
set-up is the visibility does not depend on the input state of the 
single photon, 
so we can assess how well the teleporter is working without knowing 
what is going into it. Let us see how this works.

The input for one port of the interferometer is in the arbitrary polarisation 
superposition state
\begin{equation}
\ket{\phi}_{a}=x\ket{1,0}+y\ket{0,1}
\end{equation}
where $\ket{n_{h},n_{v}}\equiv \ket{n_{h}}_{h}\otimes 
\ket{n_{v}}_{v}$, $n_{h}$ and $n_{v}$ are the photon number in the 
horizontal and vertical polarisations respectively, and 
$|x|^2+|y|^2=1$. The input of the other port is in the vacuum 
state $\ket{\phi}_{b}=\ket{0,0}$. 
The operators in the Heisenberg picture for the four input 
modes (two spatial times two polarisation) are $a_{h}$ and $a_{v}$ 
(superposition), and $b_{h}$ and $b_{v}$ (vacuum). We propagate these 
operators through the Mach-Zehnder (including the teleporter). After the first 
beamsplitter we can write
\begin{eqnarray}
c_{h,v} & = & {{1}\over{\sqrt{2}}}(a_{h,v}+b_{h,v})\nonumber\\
d_{h,v} & = & {{1}\over{\sqrt{2}}}(a_{h,v}-b_{h,v})
\end{eqnarray}
One of the beams ($c$) is then teleported. Under conditions for which 
losses can be neglected we can use Eq.~\ref{tel} to obtain
\begin{equation}
c_{h,v,T}=\lambda c_{h,v}+(\lambda 
\sqrt{H}-\sqrt{H-1})b_{h,v,1}^{\dagger}+( \sqrt{H}-\lambda 
\sqrt{H-1})b_{h,v,2}
\label{tele}
\end{equation}
The fields are recombined in phase at the final 
beamsplitter giving the outputs
\begin{eqnarray}
a_{h,v,out} & = & {{1}\over{\sqrt{2}}}(c_{h,v,T}+d_{h,v})\nonumber\\
b_{h,v,out} & = & {{1}\over{\sqrt{2}}}(c_{h,v,T}-d_{h,v})
\label{out}
\end{eqnarray}
The expectation values for photon counting at the two outputs of the 
interferometer are 
\begin{eqnarray}
<a_{out}^{\dagger}a_{out}> & = &
\bra{\phi}_{a}\bra{\phi}_{b}\bra{\phi}_{f}({a_{h,out}}^{\dagger}+
{a_{v,out}}^{\dagger})(a_{h,out}+
a_{v,out})\ket{\phi}_{a}\ket{\phi}_{b}\ket{\phi}_{f}\nonumber\\
 & = & 0.25(1+\lambda)^{2}+(\lambda
\sqrt{H}-\sqrt{H-1})^{2}\nonumber\\
<b_{out}^{\dagger}b_{out}> & = &
\bra{\phi}_{a}\bra{\phi}_{b}\bra{\phi}_{f}({b_{h,out}}^{\dagger}+
{b_{v,out}}^{\dagger})(b_{h,out}+
b_{v,out})\ket{\phi}_{a}\ket{\phi}_{b}\ket{\phi}_{f\ket{\phi}_{b}}\nonumber\\
 & = & 0.25(1-\lambda)^{2}+(\lambda \sqrt{H}-\sqrt{H-1})^{2}
\label{exp}
\end{eqnarray}
In the limit of very strong entanglement squeezing 
($V_{ent} \to 0$) we find from Eq. \ref{tele} that $c_{h,v,T}
\to c_{h,v}$ for unity gain ($\lambda=1$), 
i.e. perfect teleportation. For the same conditions (and only for 
these conditions) 
the visibility of the Mach-Zehnder outputs,

\begin{equation}
{\cal V}={{\langle a_{out}^{\dagger}a_{out}\rangle-\langle b_{out}^{\dagger}b_{out}\rangle}\over{
\langle a_{out}^{\dagger}a_{out}\rangle+\langle b_{out}^{\dagger}b_{out}\rangle}}
\label{vis}
\end{equation}
goes to one, indicating the state of the teleported arm exactly 
matches that of the unteleported arm. 
Notice that the expectation values (Eq.\ref{exp}), and 
thus the visibility, do not depend 
on the actual input state (no dependence on $x$ and $y$). Hence we 
can demonstrate that the teleporter is operating ideally even if we do 
not know the state of the input. Classical limits can be set by 
examining the visibility obtained with no entanglement ($H=1$). In 
Fig.5 we plot the visibility versus feedforward gain in the teleporter 
for the cases of no entanglement (0\%), 50\% entanglement squeezing and 90\% 
entanglement squeezing. Maximum visibility occurs for the gain 
condition
\begin{equation}
\lambda={{\sqrt{4 H-3}}\over{\sqrt{4 H+1}}}
\end{equation}
giving ${\cal V}_{max,c}=\sqrt{1/5}$ as the maximum visibility that can be 
obtained in the absence of entanglement. Increasing entanglement leads to 
increasing maximum visibility.

In the experiments we have imagined so far the level of visibility 
has been determined not only by the ability of the teleporter to 
reproduce the input polarisation 
states of the photons (the mode overlap) but also 
the efficiency with which input photons to the teleporter 
lead to correct output photons (the 
power balance). It is of interest to try to separate these effects.
We can investigate just state reproduction if we allow
attenuation to be applied to beam $d$, thus ``balancing" 
the Mach-Zehnder Interferometer by compensating for the loss introduced by the 
teleporter (see Fig.3(b)). 
The attenuated beam $d$ becomes
\begin{equation}
d_{h,v,A}=\sqrt{\eta}d_{h,v}+\sqrt{1-\eta}g_{h,v}
\end{equation}
where $g$ is another vacuum field and $\eta$ is the intensity transmission of 
the attenuator. The expectation values of the outputs 
are now
\begin{eqnarray}
\langle a_{out}^{\dagger}a_{out}\rangle & = & 0.25(\sqrt{\eta}+\lambda)^{2}+(\lambda 
\sqrt{H}-\sqrt{H-1})^{2}\nonumber\\
\langle b_{out}^{\dagger}b_{out}\rangle & = & 0.25(\sqrt{\eta}-\lambda)^{2}+
(\lambda \sqrt{H}-\sqrt{H-1})^{2}
\label{exp2}
\end{eqnarray}
In Fig.6 we plot visibility versus gain, using the attenuation $\eta$ 
to optimize the visibility ($\eta \le 1$). Now we can always achieve unit 
visibility for any finite 
level of entanglement by operating at gain
\begin{equation} 
\lambda_{opt}={{\sqrt{H-1}}\over{\sqrt{H}}}
\label{opt}
\end{equation}
and balancing the 
interferometer by setting $\eta=\lambda_{opt}^{2}$. The high 
visibility is achieved because at gain $\lambda_{opt}$ the teleporter 
behaves like pure attenuation (see Eq.~\ref{att}). That is the photon flux of the 
teleported field is reduced, but no ``spurious photons" are added to 
the field. Thus, at this gain, all output photons from the teleporter are in the 
right state, but various input photons are ``lost".

This contrast between state-reproduction and efficiency has been a 
topic of vigorous debate \cite{bra982,bou99}. It is of note that our 
interferometric test can separate the two effects. It should also be 
noted that our test is sensitive not only to the relative phase of 
the polarisation superposition, but also the overall phase of the 
teleported field. The overall phase is defined 
with respect to the field in the unteleported arm of 
the interferometer and is a constituent of the mode-overlap. If the 
overall phase is randomized by the teleporter then very low 
visibility will result from our interferometric test. At the end of 
section IV we will examine an interesting consequence of 
this additional sensitivity.

We now consider the effect of propagation loss in the two arms of the 
entangled source. Hence, referring back to Eq.\ref{ss}, we set 
$\eta_{b1}=\eta_{b2}=\eta_{b}\ne 1$. We neglect for the moment the possibility 
of internal loss in the amplification (i.e. $\eta_{a}=1$) or unequal loss 
in the two arms. With loss present (but not balancing the 
interferometer) the maximum visibility is achieved with the gain 
condition
\begin{equation}
\lambda_{max}={{\sqrt{4(H-1)+1}}\over{\sqrt{4(1-\eta_{b})+4 \eta_{b} H+1}}}
\end{equation}
In Fig.7(a) we plot maximum visibility as a function of loss for 
various levels of entanglement squeezing. Visibility is reduced quite 
rapidly. If balancing of the 
interferometer is allowed the gain condition for maximum visibility 
remains that found for no loss (Eq.\ref{opt}) but the balancing 
condition becomes $\eta=(5-4 \eta_{b})\lambda^{2}$. Once again 
visibility drops off rapidly with increasing loss (see Fig.7(b)) 
tending eventually to the classical limit as the loss 
completely wipes out the entanglement. 

The effect of loss in the amplification (or measurement stage) ($\eta_{a} 
\ne 1$)
produces very similar results to those in Fig.7, as does indeed loss 
in only the entanglement arm sent to Alice ($b_{1}$). However if loss 
is only present in the entanglement arm sent to Bob ($\eta_{b1}=1, 
\eta_{b2} \ne 1$) things are rather different. The unbalanced 
visibility is still reduced with increasing loss but when the 
interferometer is balanced one can still achieve unit visibility 
by operating at the gain 
condition
\begin{equation} 
\lambda_{opt}={{\sqrt{\eta_{b2}(H-1)}}\over{\sqrt{H}}}
\label{optn}
\end{equation}
Although the visibility is maintained the efficiency is of course 
dropping. In the limit of strong loss, $\eta_{b2} \to 0$, the 
efficiency goes to zero and no photons are teleported.

\section{More General Polarisation Input States}

So far we have assumed that the input state is a single photon number 
state. That is there is unit probability that one, and only one, 
photon arrives per measurement interval. Such states are 
yet to be demonstrated experimentally, though candidate sources have 
been proposed \cite{yam,fod}. However the results of the 
previous section don't actually rely on the input being in a number 
state. An examination of Eq.\ref{exp} shows that it is only the 
{\it expectation value} of the photon number which is important. Thus any 
input state with an average photon number of one count per measurement 
interval 
will give identical visibilities as those of the previous 
section. An example is the low photon number coherent state 
$|\phi\rangle=|\alpha_{h},\alpha_{v}\rangle$, in which 
$|\alpha_{h}|^{2}+|\alpha_{v}|^{2}=1$. 
Such a state can approximately be produced by 
strongly attenuating a stable laser beam. We 
can generalize Eq.\ref{exp} for arbitrary average input photon 
number ($\bar n$) to obtain
\begin{eqnarray}
\langle a_{out}^{\dagger}a_{out}\rangle & = & \bar n  0.25(1+\lambda)^{2}+(\lambda 
\sqrt{H}-\sqrt{H-1})^{2}\nonumber\\
\langle b_{out}^{\dagger}b_{out}\rangle & = & \bar n  0.25(1-\lambda)^{2}+
(\lambda \sqrt{H}-\sqrt{H-1})^{2}
\label{exp3}
\end{eqnarray}
Maximum visibility now occurs for the gain 
condition
\begin{equation}
\lambda_{max}={{\sqrt{4 H+\bar n-4}}\over{\sqrt{4 H+\bar n}}}
\end{equation}
giving ${\cal V}_{max,c}=\sqrt{\bar n/(\bar n+4)}$ for the maximum
classical visibility. As might be expected higher maximum visibilities can be 
achieved with only a classical channel as the average photon number 
increases and the input becomes more like a classical field. For 
average photon numbers less than one the maximum achievable 
visibility is reduced. This is basically a signal to noise effect. The 
penalty in classical teleportation arises from amplification of vacuum 
fluctuations ($v_{1}$) introduced in the ``measurement'' process. For 
low photon numbers this noise is large compared to the signal leading 
to low visibility. For large photon numbers the noise can become 
negligible compared to the signal leading to high visibilities. Fig.8 
illustrates the change in $\lambda_{max}$ and $V_{max}$ as a function 
of entanglement for various values of the input photon number.

Single photon number states can be realized conditionally by using 
number entangled states. It is instructive to investigate 
this special case (see Fig.4(b)). A low efficiency, 
non-degenerate parametric amplifier (down converter) 
can produce pairs of photons in the polarisation-number entangled state 
\begin{equation}
|\phi\rangle_{a,a'}\approx 
|0,0\rangle_{a}|0,0\rangle_{a'}+\chi(|1,0\rangle_{a}|1,0\rangle_{a'}+
|0,1\rangle_{a}|0,1\rangle_{a'})
\end{equation}
where $a$ and $a'$ are the two, spatially separated fields and 
$\chi$ is the conversion efficiency. We have assumed $\chi<<1$ and 
neglected higher order terms in $\chi$. As 
before $a$ is the input field to the interferometer plus teleporter and is 
transformed as per Eq.\ref{out}. We can either analyze the raw 
visibility of the outputs or the conditional visibility. Beam $a$ by 
itself is in the unpolarised mixed state, given by the reduced density 
operator
\begin{equation}
\rho_{a}\approx |0,0\rangle\langle 0,0|+\chi^{2}(|0,1\rangle\langle 0,1|+
|1,0\rangle\langle 1,0|)
\end{equation}
The raw count rates are thus calculated using $\langle a^{\dagger}a\rangle=Tr[\rho 
a^{\dagger}a]$. As would be expected the raw visibility is as predicted by 
Eq.\ref{exp3} with $\bar n=\chi^{2}$. Because $\chi$ is small, 
classical teleportation visibilities will be low. However with 
teleportation entanglement they can, in principle, reach unity. It is 
important to note that the commonly used measure of teleportation, 
fidelity, cannot be used to judge teleportation of such a mixed state 
\cite{note}. 
The fidelity between mixed input and output states is defined by 
\cite{bar96}
\begin{equation}
F=Tr[ \sqrt{\rho_{a}^{1/2}\rho_{out}\rho_{a}^{1/2}}]
\end{equation}
If $\rho_{a}=\rho_{out}$ then $F=1$. But this can easily be arranged 
by a cheating Alice and Bob without using entanglement. This is because {\it 
any} unpolarised 
mixed state with average photon number $\chi^{2}$ will have a density 
operator equal to $\rho_{a}$. Only by making measurements of the 
joint state of $a$ and $a'$ before and after the teleporter and 
calculating a global fidelity can a high fidelity be considered proof 
of quantum teleportation. In contrast a local interferometric test on only 
$a$ unambiguously judges the quality of the 
teleporter. This is due to the sensitivity of the teleporter to the 
overall phase of the field. As a result high visibilities are only 
possible when Alice and Bob share entanglement.
 
Conditional visibilities can be obtained by making the coincidence 
counts 
$\langle \phi|_{a,a'}\langle \phi|_{b} a'^{\dagger}a'a_{out}^{\dagger}
a_{out}|\phi\rangle_{b}|\phi\rangle_{a,a'}$ and 
$\langle \phi|_{a,a'}\langle \phi|_{b} a'^{\dagger}a'b_{out}^{\dagger}
b_{out}|\phi\rangle_{b}|\phi\rangle_{a,a'}$. Now counts are only recorded if 
a photon has simultaneously been detected in beam $a'$. This guarantees 
that only counts corresponding to times when a photon is launched into the 
interferometer are recorded. The visibilities then correspond to 
those obtained in section 2 with single photon input states. This 
result is conceptually different from the case of an average of one photon 
per measurement interval because it can be arranged, 
to a high probability, that only 
one photon is ever present at one time in the interferometer. 

\section{Continuous Variable Inputs}

We now consider a very different type of input state and detection 
technique. Our input beam will now potentially be a ``bright'' beam. 
However our interest will centre only on the state of the ``side-bands'' 
of the beam at some RF frequencies $\pm \omega$ around the central 
frequency. We will require that 
$\omega$ is sufficiently large that the power in the 
side-bands at that frequency are of the order of one photon per 
second. 
Typically, for solid-state lasers, $\omega\rangle10Mhz$ will suffice. 
Instead of considering the polarisation state of the light, as in the 
previous sections, we will now consider the field state of the 
side-bands, as characterized by their distribution of power between phase and 
amplitude fluctuations. The total power in the side-bands at the 
outputs can be 
measured using optical homodyne techniques and visibilities 
constructed. These visibilities behave identically 
to those in the photon counting case provided the average photon 
number in the sidebands is equal to $\bar n$. This is quite surprising 
given the incompleteness of the formal analogy between single photon 
polarisation states and single mode continuous variable states.

The proposed set-up is shown in Fig.4(c). It is identical to that for the 
single photon input except for the homodyne detection systems at the 
outputs instead of photon counters. The output beams are divided in 
half at beamsplitters and sent to homodyne detectors which detect 
orthogonal quadrature amplitudes, i.e.
\begin{eqnarray}
X^{+}(\omega) & = & e^{i \theta}a(\omega)+e^{-i 
\theta}a^{\dagger}(\omega)\nonumber\\
X^{-}(\omega) & = & e^{i (\theta+\pi/2)}a(\omega)+e^{-i 
(\theta+\pi/2)}a^{\dagger}(\omega)
\end{eqnarray}
where the absolute quadrature angle, $\theta$, is arbitrary. Although 
the homodyne detection itself can be ideal, the splitting of the beams 
at the beamsplitters inevitably introduces vacuum noise (this must 
occur because orthogonal quadratures constitute conjugate 
observables). Thus the detection results are
\begin{eqnarray}
X_{a}^{+}(\omega) & = & {{1}\over{\sqrt{2}}}(a_{out}(\omega)+
a_{out}^{\dagger}(\omega)+v_{d1}+v_{d1}^{\dagger})\nonumber\\
X_{a}^{-}(\omega) & = & {{i}\over{\sqrt{2}}} (a_{out}^{\dagger}(\omega)- 
a_{out}(\omega)
+v_{d1}-v_{d1}^{\dagger})\nonumber\\
X_{b}^{+}(\omega) & = & {{1}\over{\sqrt{2}}}(b_{out}(\omega)+
b_{out}^{\dagger}(\omega)+v_{d2}+v_{d2}^{\dagger})\nonumber\\
X_{b}^{-}(\omega) & = & {{i}\over{\sqrt{2}}} (b_{out}^{\dagger}(\omega)-
b_{out}(\omega)
+v_{d2}-v_{d2}^{\dagger})
\end{eqnarray}
where the arbitrary angle, $\theta$ has been set to zero for simplicity.
The penalty vacuum noise is represented as usual by $v$'s. Consider 
adding the 
photocurrents from each beam with a $\pi/2$ phase shift. 
This could be achieved by imposing a delay of $\tau$ to one of the currents 
such that $\tau \omega= \pi/2$. This gives photocurrents
\begin{eqnarray}
A(\omega) & = & X_{a}^{+}+i 
X_{a}^{-}=\sqrt{2}(a_{out}+v_{d1}^{\dagger})\nonumber\\
B(\omega) & = & X_{b}^{+}+i X_{b}^{-}=\sqrt{2}(b_{out}+v_{d2}^{\dagger})
\end{eqnarray}
These photocurrents could then be fed into spectrum analyzers which give 
the photon number spectra
\begin{eqnarray}
V_{A}(\omega) & = & \langle |X_{a}^{+}+i X_{a}^{-}|^{2}\rangle=2 
\langle a_{out}^{\dagger}(\omega)a_{out}(\omega)\rangle+2\nonumber\\
V_{B}(\omega) & = & \langle |X_{b}^{+}+i X_{b}^{-}|^{2}\rangle=2 
\langle b_{out}^{\dagger}(\omega)b_{out}(\omega)\rangle+2
\end{eqnarray}
We can then define, in analogy with the photon counting case 
(Eq.\ref{vis}),the spectral visibility as
\begin{eqnarray}
{\cal V} & = & {{\langle a_{out}^{\dagger}(\omega)a_{out}(\omega)\rangle-
\langle b_{out}^{\dagger}(\omega)b_{out}(\omega)\rangle}
\over{\langle a_{out}^{\dagger}(\omega)a_{out}(\omega)\rangle+
\langle b_{out}^{\dagger}(\omega)b_{out}(\omega)\rangle}}\nonumber\\
& = & {{V_{A}-V_{B}}\over{V_{A}+V_{B}-4}}
\end{eqnarray}
Note that for an arbitrary field we can also write
\begin{equation}
V_{In}(\omega) =\langle |X^{+}+i X^{-}|^{2}\rangle=V^{+}+V^{-}-2
\end{equation}
where $V^{+}=\langle |X^{+}|^{2}\rangle$ and $V^{-}=\langle |X^{-}|^{2}\rangle$. 
Hence we can 
make the identification
\begin{equation}
\langle a_{in}^{\dagger}(\omega)a_{in}(\omega)\rangle=
{{1}\over{4}}(V^{+}+V^{-})-{{1}\over{2}}
\label{id}
\end{equation}
Eq.\ref{id} allows us to construct visibilities directly from 
individually measured 
orthogonal quadrature spectral variances. Also it allows us to compare 
the visibilities obtained here with those of the previous sections.
In order to make such comparisons with the photon counting visibilities 
we observe 
that $\langle a_{in}^{\dagger}(\omega)a_{in}(\omega)\rangle$ is the photon number 
in the upper frequency component of the field only. Thus the total 
average photon number of upper and lower side-bands (assuming a 
frequency symmetric input state) is $\bar n(\pm \omega)=2 
\langle a_{in}^{\dagger}(\omega)a_{in}(\omega)\rangle$. This is 
similar to the summing of the average photon numbers for both 
polarization modes in the discrete case. For equivalent average 
photon numbers (Eq.\ref{exp3} with $\bar n(\pm \omega)\equiv \bar n$) 
{\it all} the predictions of the low photon number
visibilities are exactly reproduced in the continuous variable case, 
including the ability to re-balance the interferometer and obtain unit 
visibilities. 

The preceding analysis has shown that interferometric tests of 
quantum teleportation for
unknown continuous variable states of a fixed average photon number 
can also be performed.
Let us consider a couple of examples. For an arbitrary input field 
there will be some particular value of 
$\theta$ for which the conjugate spectral variances reach maximum and 
minimum values, $V_{max}^{+}$ and $V_{min}^{-}$ respectively. A 
minimum uncertainty state obeys the equality 
$V_{max}^{+}V_{min}^{-}=1$. It is convenient to discuss our examples 
in terms of these quadratures. 
Suppose our input field is quantum noise limited but with a 
small classical signal imposed at an arbitrary quadrature angle. This 
is equivalent to a coherent state of a particular amplitude but 
unknown phase. For this input $V_{max}^{+}=V_{s}+1$ and 
$V_{min}^{-}=1$, where $V_{s}$ is the signal power. If $V_{s}=2$ 
then spectral visibilities identical to the single photon counting 
visibilities will be observed. Alternatively the input state may be 
squeezed at some arbitrary angle such that $V_{max}^{+}>1>V_{min}^{-}$. 
If $V_{max}^{+}=1/(2-\sqrt{3})$ and $V_{min}^{-}=(2-\sqrt{3})$ then 
again spectral visibilities will be identical to the single photon counting 
visibilities.

These results are significant as reliable teleportation of spectral 
components is technologically less challenging than single photon 
experiments and are thus likely to form a significant part of future 
quantum information research.

\section{Conclusion}

We have examined an interferometric test of the efficacy of teleportation. 
Though more specific than other teleportation figures of merit, 
interferometric visibility is clearly of importance in applications of 
teleportation in quantum information processing. Unique 
characteristics of this arrangement are: (i) it  
doesn't require the tester to know the input state of the light, only 
the average power; (ii) the ability of the teleporter to reconstruct 
both the relative and global phase of the field is tested directly; 
and (iii) one can directly
test the state reconstruction ability of the teleporter separately 
from or together with its efficiency.

The teleportation efficacy is characterized by the visibility 
between the two outputs of a Mach-Zehnder interferometer when the 
teleporter to be tested is placed in one of the arms. 
We have contrasted the results obtained with no entanglement and varying 
levels of squeezing entanglement using continuous 
variable teleportation. A clear classical limit (i.e. with no 
entanglement) to the
visibility was demonstrated and its dependence on input average 
photon number investigated. For an average photon count of one per 
measurement interval the classical limit was ${\cal V} \le \sqrt{1/5}$. 
Higher classical visibilities could be obtained with greater photon 
flux. The classical limit was lower with smaller photon flux. High 
visibilities (close to one) could only be obtained (for low photon 
flux) with high levels of entanglement and low levels of loss. These 
are the requirements for high efficiency teleportation. However decreased 
photon flux in the teleported arm (reduced efficiency) can be compensated 
by re-balancing 
the unteleported arm of the interferometer. In this way state reconstruction 
can be tested separately from efficiency. We find that, provided losses 
are small, ideal state reconstruction can be achieved for any level 
of entanglement squeezing. This is characterized by unit 
visibility in the balanced interferometer 
with finite levels of entanglement. Losses 
reduce visibilities but the general trends remain the 
same. 

A generalization of the technique to continuous variable inputs 
was presented. With suitable interpretation it was found that 
the visibilities exhibited identical behavior to their discrete 
variable counterparts.

We believe that tests of the kind outlined in this 
paper will be essential if quantum teleportation is to be incorporated in 
reliable quantum information networks.

We wish to thank A.G.White and G.J.Milburn for helpful discussions. 
This work was supported by the Australian Research Council.

\section*{Appendix}

Some readers may find it unusual that the classical channel $a_{c}$ is 
described by an operator. This is a standard feature of the treatment 
of classical channels in the Heisenberg picture, {\it 
not} a consequence of 
our particular choice of an optical classical channel or our 
particular choice of teleporter model. The different treatments of 
classical channels between the Heisenberg and Schroedinger pictures 
are contrasted for quantum limited feedback in Ref.\cite{wis94}.
That $a_{c}$ is truly a classical channel can be demonstrated easily 
via the no-cloning theorem \cite{not1} which states that a quantum 
system can not be duplicated without penalty. If the quantum nature 
of $a_{c}$ is significant in the teleportation process then the
no-cloning theorem would predict that duplication of $a_{c}$ would 
lead to a significant degradation in the quality of the teleported 
state. An optimum continuous 
variable cloner can be constructed from the combination of a linear 
amplifier of gain 2 followed by a 50:50 beamsplitter. Applying this 
to $a_{c}$ produces the two clones $a_{c}'$ and $a_{c}''$ given by
\begin{eqnarray}
	a_{c}' & = & a_{c}+{{1}\over{\sqrt{2}}}(v_{c1}^{\dagger}+v_{c2})\nonumber\\
	a_{c}''& = & a_{c}+{{1}\over{\sqrt{2}}}(v_{c3}^{\dagger}-v_{c2})
\end{eqnarray}
where the $v$'s are vacuum modes. Suppose Bob uses $a_{c}'$ for the 
reconstruction. He will produce the output
\begin{equation}
a_{out} = \lambda  a_{in}+
(\lambda  \sqrt{H}-\sqrt{H-1})v_{3}^{\dagger}+
(\sqrt{H}-\lambda \sqrt{H-1}) v_{4}+\sqrt{\varepsilon}
{{1}\over{\sqrt{2}}}(v_{c1}^{\dagger}+v_{c2})
\label{telap}
\end{equation}
The final term is due to the cloning process. But in the classical 
channel limit we have $\varepsilon \to 0$ and hence this final term can be 
neglected and Eq.\ref{tel} reduces to Eq.\ref{telap}. Arbitrarily good 
reconstruction of the input beam is still possible. A same result 
holds if Bob were to use the other clone, $a_{c}''$, for the 
reconstruction. Thus the cloning procedure does not change the 
quantum properties of the output and so $a_{c}$ must be considered a 
classical channel.

\begin{figure}
 \caption{Schematics of all optical teleporter. In (a) a classical 
 teleporter is shown (i.e. with no entanglement). In (b) the 
 inclusion of entanglement (EPR) is shown. In (c) the separate 
 teleportation of the two polarisation modes is represented. TV and 
 TH are the teleporters for the vertical and horizontal polarisation 
 components respectively. PBS stands for polarising beam-splitter.}
\end{figure}

\begin{figure}
 \caption{Schematic of the linear amplifier used in the teleporters. 
 The PA's stand for parametric amplifiers which are pumped in phase 
 ($E$) and out of phase ($-E$) with the field. }
\end{figure}

\begin{figure}
\caption{Schematics of interferometric test arrangements}
\end{figure}

\begin{figure}
\caption{Schematics of different input state-measurement techniques.}
\end{figure}

\begin{figure}
 \caption{Visibility versus gain for the set-up shown in Fig.3(a) and 
 various levels of entanglement (0\%, 50\% and 90\%).}
\end{figure}

\begin{figure}
 \caption{Visibility versus gain with ``attenuation balancing'' 
 (set-up shown in Fig.3(b)) for various levels of 
 entanglement (0\%, 50\% and 90\%).}
\end{figure}

\begin{figure}
 \caption{The effect of loss on the visibility. In (a) the maximum 
 visibility is plotted versus the transmission efficiency of the 
 entangled beams for various levels of 
 entanglement (0\%, 50\% and 90\%). In (b) balancing of the 
 interferometer is allowed (plot is for 50\% entanglement)}
\end{figure}

\begin{figure}
\caption{Gain for maximum visibility ($\lambda_{max}^{2}$) and maximum 
visibility thus achieved ($V_{max}$) versus level of entanglement for 
various average input photon numbers ($\bar n=0.25, 1.0, 4.0$).}
\end{figure}


\begin{thebibliography}{99}
\bibitem{note} In principle the criteria proposed in \cite{ral98} can 
be used for mixed states.
\end{thebibliography}
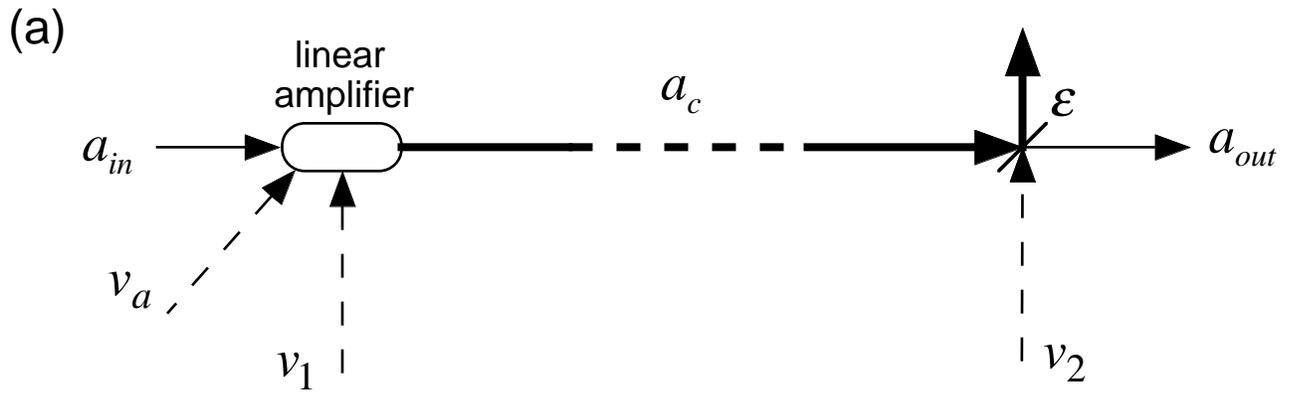
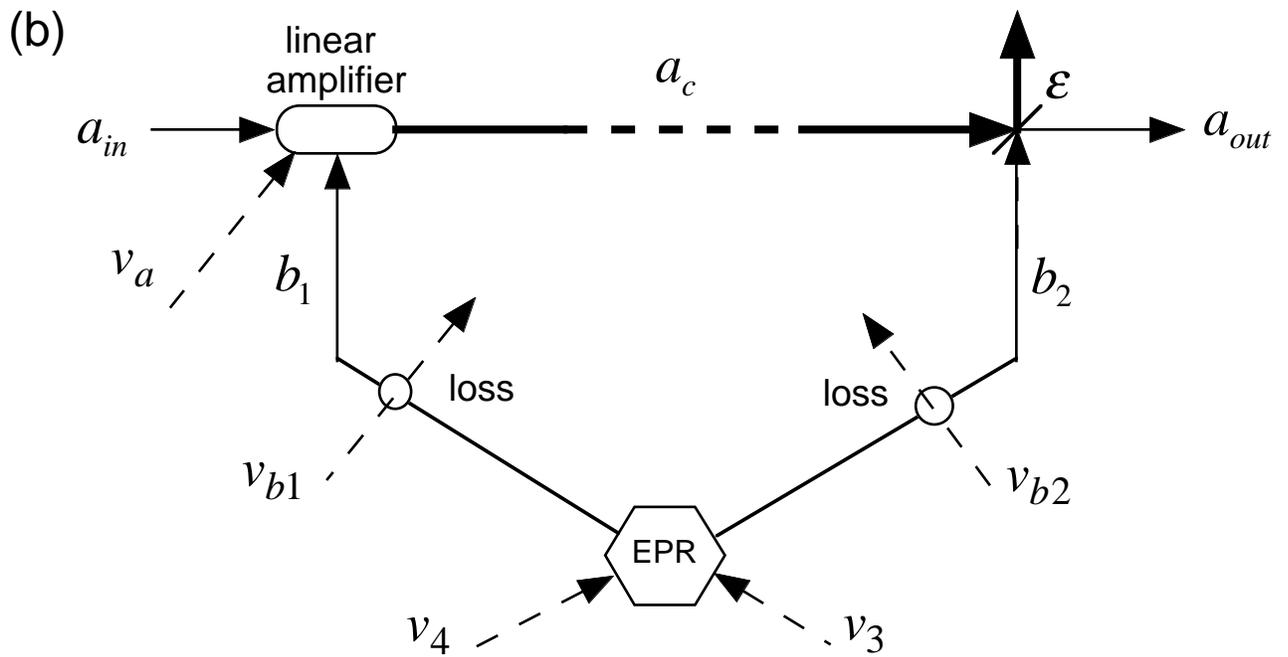
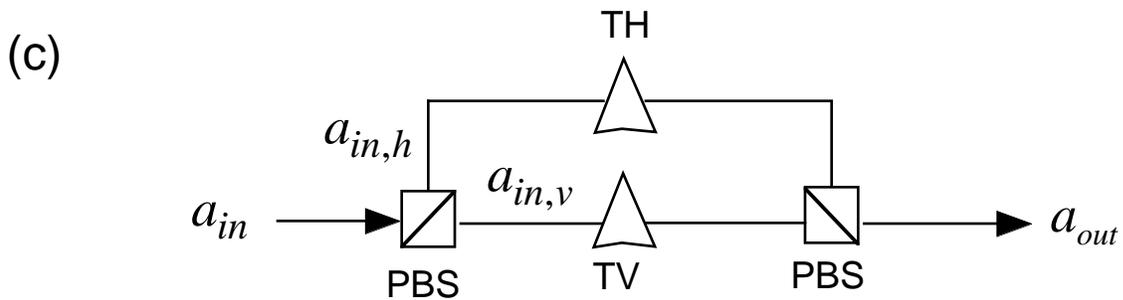
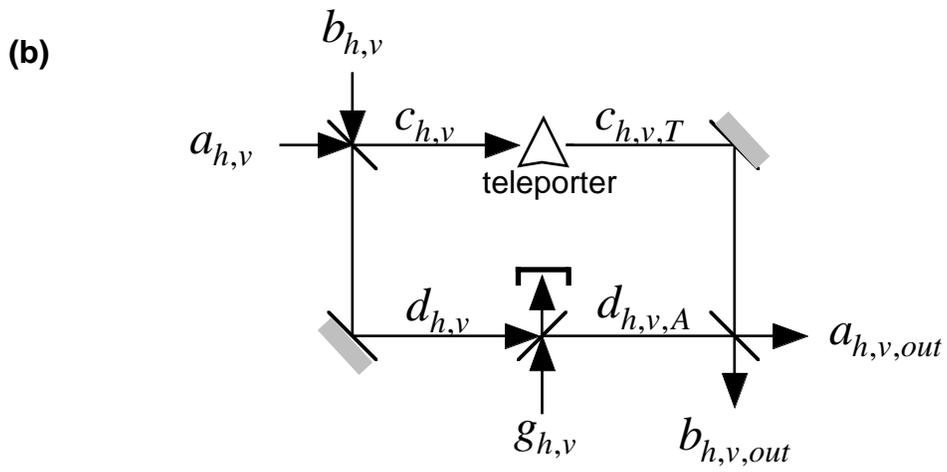
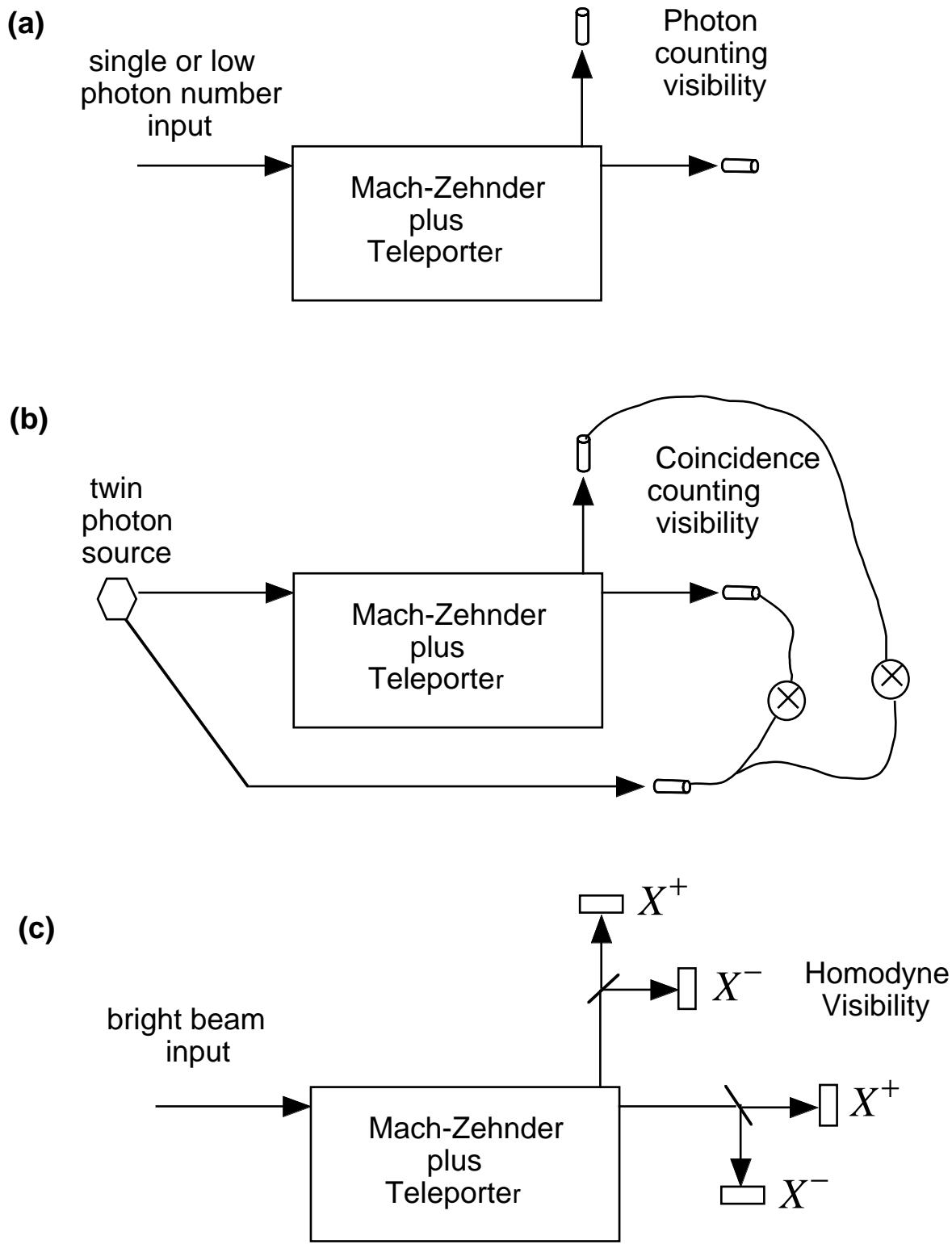
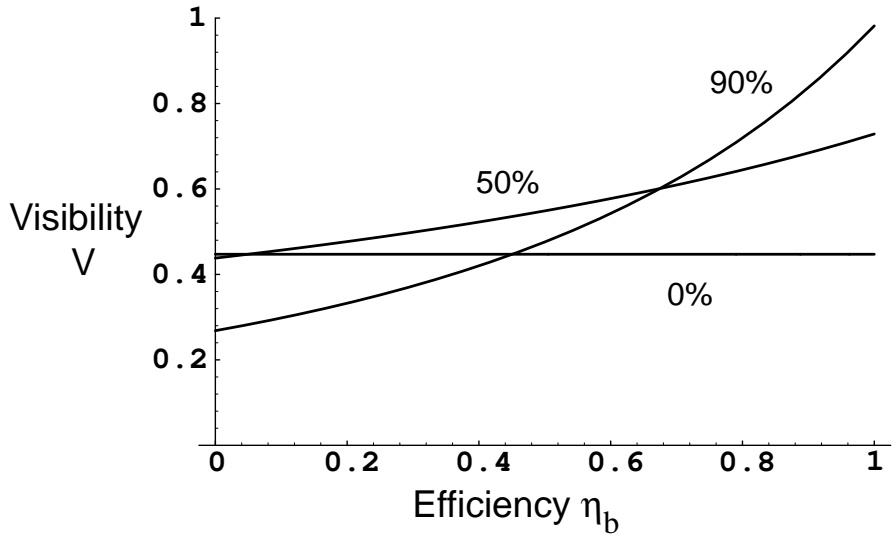
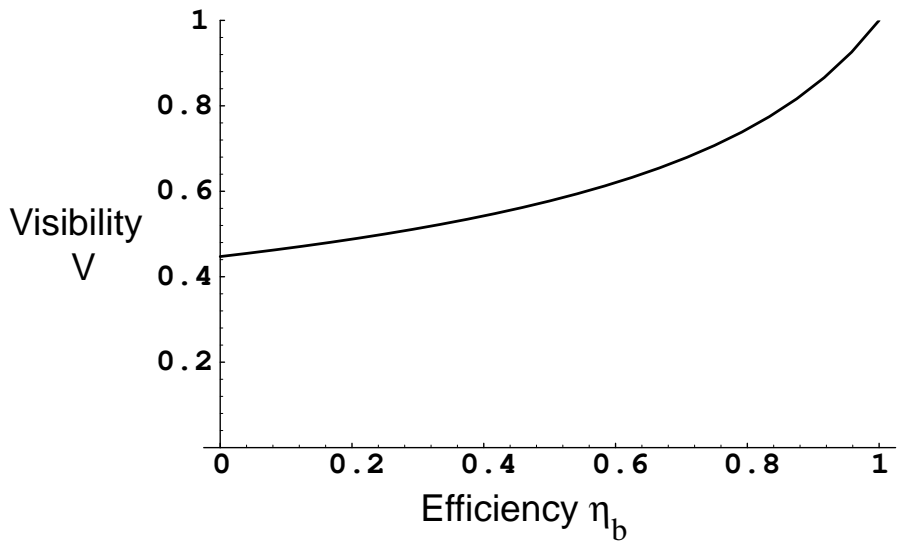
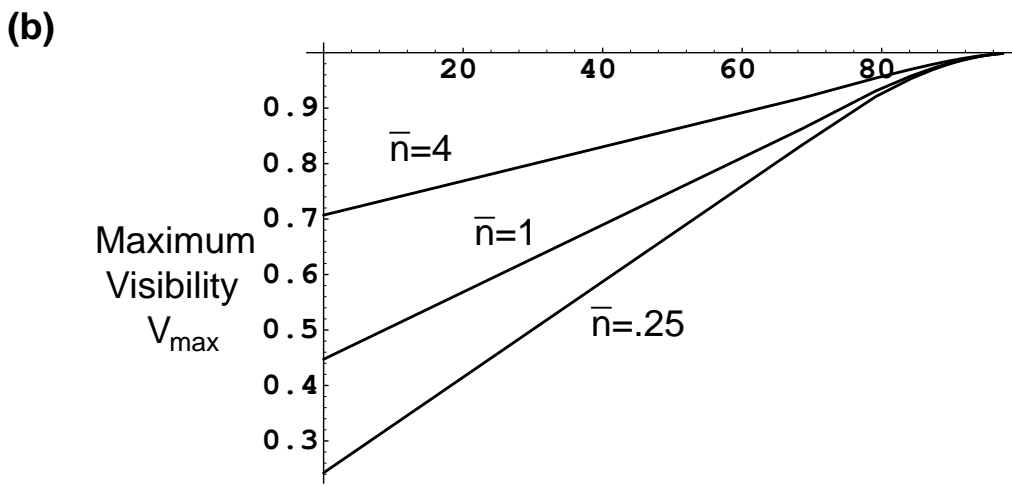
\end{document}